\documentstyle[12pt]{article}
\begin{document}
\renewcommand{\theequation}{\thesection .\arabic{equation}}
\newcommand{\beq}{\begin{equation}}
\newcommand{\eeq}{\end{equation}}
\newcommand{\beqn}{\begin{eqnarray}}
\newcommand{\eeqn}{\end{eqnarray}}
\newcommand{\slp}{\raise.15ex\hbox{$/$}\kern-.57em\hbox{$ \partial $}}
\newcommand{\lnA}{\raise.15ex\hbox{$/$}\kern-.57em\hbox{$A$}}
\newcommand{\bP}{\bar{\Psi}}
\newcommand{\bc}{\bar{\chi}}
\newcommand{\hs}{\hspace*{0.6cm}}

\title{Path-integral formulation of backward and umklapp scattering for 1d spinless fermions}
\author{V.I. Fern\'andez$^{a,b}$ and C.M.Na\'on$^{a,b}$}
\date{October 1998}
\maketitle

\def\thepage{\protect\raisebox{0ex}{\ } La Plata Th 98-15}
\thispagestyle{headings}
\markright{\thepage}

\begin{abstract}

\hs We present a (1+1)-dimensional fermionic QFT with non-local couplings between currents.
This model describes an ensemble of spinless fermions interacting through forward, backward
and umklapp scattering processes.
We express the vacuum to vacuum functional in terms of a non trivial fermionic
determinant. Using path-integral methods we find a bosonic representation for this
determinant. Thus we obtain an effective action depending on three scalar
fields, two of which correspond to the physical collective excitations whereas the third one
is an auxiliary field that is left to be integrated by means of an  approximate technique.
\end{abstract}

\vspace{3cm}
Pacs: 05.30.Fk\\
\hspace*{1,7 cm}11.10.Lm\\
\hspace*{1,7 cm}71.10.Pm

\noindent --------------------------------

\noindent $^a$ {\footnotesize Depto. de F\'{\i}sica.  Universidad
Nacional de La Plata.  CC 67, 1900 La Plata, Argentina.}

\noindent $^b$ {\footnotesize Consejo Nacional de Investigaciones
Cient\'{\i}ficas y T\'ecnicas, Argentina.\\
e-mail: naon@venus.fisica.unlp.edu.ar}

\newpage
\pagenumbering{arabic}

\section{Introduction}

\hs In recent years there has been a renewed interest in the study of low-dimensional field
theories. In particular, research on the one-dimensional (1d)
fermionic gas has been very active, mainly due to the actual
fabrication of the so called quantum wires \cite{Voit}. One of the most interesting aspects
of these systems is the possibility of having a deviation from the usual
Fermi-liquid behavior. This phenomenon was systematically examined by Haldane
\cite{Haldane} who coined the term Luttinger-liquid behavior to name this
new physical situation in which the Fermi surface disappears and the spectrum
contains only collective modes. Perhaps the simplest theoretical framework
that presents this feature is the Tomonaga-Luttinger (TL) model \cite{TL},
a many-body system of right and left-moving particles interacting through their charge
densities. In a recent series of papers \cite{NLT} an alternative, field-theoretical
approach was developed to consider this problem. In these works a non-local and non-covariant version of the Thirring model \cite{Thirring} \cite{Klaiber} was introduced, in which the
fermionic densities and currents are coupled through
bilocal, distance-dependent potentials. This non-local Thirring model (NLT)
contains the TL model as a particular case. Although it constitutes an elegant framework to
analyze the 1d many-body problem, one seriuos limitation appears if one tries to make
contact with quantum wires phenomenology. Indeed, one has to recall that, from a
perturbative point of view, the building blocks of the NLT are the forward-scattering (fs)
processes which are supposed to dominate the scene only in the low transferred momentum
limit.
This means that in its present form it can only provide a very crude description of the
Luttinger liquid equilibrium and transport properties. The main goal of this paper is to
start developing an improved version of the NLT in which larger momentum transfers are taken
into account. Although the most interesting effects of these contributions, such as the
occurrence of mass-gaps in the normal modes spectra, are known to take place for
spin-$\frac{1}{2}$ particles, here, for illustrative purposes we shall focus our attention
on the spinless case.  In Section 2 we show that the manipulations used to write the vacuum
functional of the fs problem in terms of a fermionic determinant, also work for the NLT,
even when backward (bs) \cite{back} and umklapp (us) \cite{um} scattering are considered.
We get a quite peculiar fermionic determinant which resembles the one obtained in the path-integral study \cite{N} of massive fermions in (1+1) dimensions \cite{massive}.
In Section 3 we obtain a bosonic representation for  this non-trivial determinant.
This, in turn, allowed us to find a completely bosonized effective action for the model
under consideration. Finally, in Section 4, we summarize the
main points of our investigation.

\section{The Model}
\setcounter{equation}{0}

\hs In this section we begin the study of an extended version of the NLT which includes the
contribution of forward, backward and umklapp scattering. Following the formulation proposed
in \cite{NLT} we shall attempt to describe these interactions by means of a fermionic QFT
with Euclidean action given by

\beq
S = S_0 + S_{fs} + S_{bs} + S_{us}
\label{1}
\eeq
\noindent where

\beq
S_0 = \int d^2x~ \bP i\slp\Psi
\label{2}
\eeq

\noindent is the unperturbed action associated to a linearized free dispersion relation.
The contributions of the different scattering processes will be written as

\beq
S_{fs} = - \frac{g^2}{2} \int d^2 x ~d^2 y~
( \bP \gamma_{\mu} \Psi ) (x)~ V_{(\mu)}(x,y) ~( \bP \gamma_{\mu} \Psi ) (y)
\label{3}
\eeq
\noindent and
\beqn
S_{bs} + S_{us} &=& -\frac{g'^2}{2} \int d^2x~ d^2y  (\bP~\Gamma_{\mu} \Psi)(x)~U_{(\mu)}(x,y)(\bP~\Gamma_{\mu} \Psi)(y)
\label{4}
\eeqn

\noindent where the $\gamma_{\mu}'s$ are the usual two-dimensional Dirac matrices and
$\Gamma_{0}=1$, $\Gamma_{1}=\gamma_{5}$. \noindent Please keep in mind that no sum over
repeated indices is implied when a subindex $(\mu)$ is involved.
Let us also mention that the coupling potentials
$V_{(\mu)}$ and $U_{(\mu)}$ are assumed to depend on the distance $\mid x-y \mid$ and can be
expressed in terms of Solyom's "g-ology" \cite{Solyom} as
\beqn
V_{(0)}(x,y) &=& \frac{1}{g^2}(g_2 +g_4)(x,y) \nonumber\\
V_{(1)}(x,y) &=& \frac{1}{g^2}(g_2 -g_4)(x,y)
\label{5}
\eeqn
\beqn
U_{(0)}(x,y) &=& \frac{1}{g'^2}(g_3+g_1)(x,y) \nonumber\\
U_{(1)}(x,y) &=& \frac{1}{g'^2}(g_3 -g_1)(x,y)
\label{6}
\eeqn
\noindent In the above equations $g$ and $g'$ are just numerical constants that could be set equal to one. We keep them to facilitate comparison of our results with those corresponding to the usual Thirring model. Indeed, this case is obtained by choosing $g'=0$ and $V_{(0)}(x,y)=V_{(1)}(x,y)=\delta^2(x-y)$. On the other hand, the non-covariant limit $g'=0$, $V_{(1)}(x,y)=0$ gives the TL model \cite{TL}.

\hs The terms in the action containing $g_2$ and $g_4$ represent forward scattering events,
in which the associated momentum transfer is small. In the $g_2$ processes the two branches
(left and right-moving particles) are coupled, whereas in the $g_4$ processes all four
participating electrons belong to the same branch. On the other hand $g_1$ and $g_3$ are
related to scattering diagrams with larger momentum transfers of the order of $2 k_F$ (bs)
and $4 k_F$ (us) respectively (This last contribution is important only if the band is
half-filled).
\hs Let us now turn to the treatment of the partition function. At this point we recall that
in ref.\cite{NLT} we wrote the fs piece of the action in a localized way :
\beq
S_{fs} = -\frac{g^2}{2} \int d^2x~  J_{\mu}  K_{\mu}.
\label{7}
\eeq

\noindent where $J_{\mu}$ is the usual fermionic current, and $ K_{\mu}$ is a new current defined as

\beq
 K_{\mu}(x) = \int d^2y~ V_{(\mu)}(x,y)  J_{\mu}(y).
\label{8}
\eeq
Using a functional delta and introducing auxiliary bosonic fields in the path-integral representation of the partition function $Z$ we were able to write (see \cite{NLT} for details):
\beq
Z = N \int \mbox{$\mathcal{D}$}\bP \mbox{$\mathcal{D}$}\Psi \mbox{$\mathcal{D}$}\tilde{A}_{\mu}\mbox{$\mathcal{D}$}\tilde{B}_{\mu}~ exp\{-\int
d^2x [\bP i\slp\Psi +\tilde{A}_{\mu}\tilde{B}_{\mu}
+ \frac{g}{\sqrt{2}}(\tilde{A}_{\mu}
J_{\mu} + \tilde{B}_{\mu}K_{\mu})] \}
\label{9}
\eeq

\noindent If we define
\beq
\bar{B}_{\mu}(x) = \int d^2y~ V_{(\mu)}(y,x)\tilde{B}_{\mu}(y),
\label{10}
\eeq
\beq
\tilde{B}_{\mu}(x) = \int d^2y~ b_{(\mu)}(y,x) \bar{B}_{\mu}(y),
\label{11}
\eeq

\noindent with $b_{(\mu)}(y,x)$ satisfying

\beq
\int d^2y~ b_{(\mu)}(y,x) V_{(\mu)}(z,y) = \delta^2 (x-z),
\label{12}
\eeq

\noindent and change auxiliary variables in the form

\beq
A_{\mu}=\frac{1}{\sqrt{2}}(\tilde{A}_{\mu} +\bar{B}_{\mu}),
\label{13}
\eeq

\beq
B_{\mu}=\frac{1}{\sqrt{2}}(\tilde{A}_{\mu} - \bar{B}_{\mu}),
\label{14}
\eeq

\noindent we obtain
\beq
Z = N \int \mbox{$\mathcal{D}$}\bP \mbox{$\mathcal{D}$}\Psi \mbox{$\mathcal{D}$}{A}_{\mu}\mbox{$\mathcal{D}$}{B}_{\mu}~e^{-S(A,B)-S_{bs}-S_{us}}~
 exp\{-\int
d^2x ~ \bP (i\slp- g \lnA)\Psi \}
\label{15}
\eeq

\noindent where
\beqn
S(A,B)=\frac{1}{2}\int d^2x~ d^2y ~b_{(\mu)}(x,y)[A_{\mu}(x)
      A_{\mu}(y)-B_{\mu}(x)B_{\mu}(y)]
\label{16}
\eeqn

\noindent The Jacobian associated with the change $(\tilde{A}, \tilde
{B})\rightarrow (A,B)$ is field- independent and can then be absorbed in
the normalization constant $N$. Moreover, we see
that the $B$-field is completely decoupled from both the $A$-field and the
fermion field. Keeping this in mind, it is instructive to try to recover
the partition function corresponding to the usual covariant Thirring model
($b_{(0)}(y,x)$ =$ b_{(1)}(x,y)$ =$ \delta^2(x-y)$, and $g'=0$), starting from (\ref{15}).
In doing so one readily discovers that $B_{\mu}$ describes a negative-metric
state
whose contribution must be factorized and absorbed in $N$ in order to
get a sensible answer for Z. This procedure paralells, in the path-integral
framework, the operator approach of Klaiber \cite{Klaiber}, which
precludes the use of an indefinite-metric Hilbert space. Consequently, from now on we shall only consider the A contribution.

At this stage we see that when bs and us processes are disregarded the procedure we have just sketched allows us to express $Z$
in terms of a fermionic determinant. Now we will show that this goal can also be achieved when the larger momentum transfers are taken into account.
To this end we write:

\beq
S_{bs} + S_{us} = -\frac{g'^2}{2} \int d^2x~  L_{\mu}  M_{\mu}
\label{17}
\eeq

\noindent where $L_{\mu}$ and $M_{\mu}$ are fermionic bilinears defined as

\beq
L_{\mu}(x) = \bP(x)  \Gamma_{\mu} \Psi(x),
\label{18}
\eeq

\beq
 M_{\mu}(x) = \int d^2y~ U_{(\mu)}(x,y)  L_{\mu}(y).
\label{19}
\eeq

\noindent Thus it is evident that we can follow the same prescriptions as above, with $L_{\mu}$ and $M_{\mu}$ playing the same roles as $J_{\mu}$ and $K_{\mu}$, respectively. After the elimination of a new negative metric state whose decoupled partition function is absorbed in the normalization factor, as before, one obtains
\beq
Z = N \int  \mbox{$\mathcal{D}$} A_{\mu}   \mbox{$\mathcal{D}$} C_{\mu}~ \det (i \slp - g \lnA - g' \Gamma _{\mu}C_{\mu}) ~ \exp \{ -S[A] - S[C] \}
\label{20}
\eeq

\noindent where

\beqn
S[A_{\mu}] = \frac{1}{2}\int d^2 x d^2 y ~ A_{\mu}(x) b_{(\mu)}A_{\mu}(y)  \nonumber \\
S[C_{\mu}] = \frac{1}{2}\int d^2 x d^2 y ~ C_{\mu}(x) d_{(\mu)}C_{\mu}(y)  \nonumber \\
\label{21}
\eeqn

\noindent and
\beq
\int d^2y~ d_{(\mu)}(y,x) U_{(\mu)}(z,y) = \delta^2 (x-z),
\label{22}
\eeq

This is our first interesting result: we have been able to express $Z$
in terms of a fermionic determinant. Let us stress, however, that this determinant is a highly non trivial one. Indeed, the term in $g'$ is not only a massive-like term (in the sense that it is diagonal in the Dirac matrices space) but it also depends on the auxiliary field $C_{\mu}(x)$. In the next section we shall show how to deal with this determinant.

\section{Boson representation for the determinant}
\setcounter{equation}{0}

\noindent In this section we shall combine a chiral change in the fermionic path-integral measure with a formal expansion in $g'$ in order to get a bosonic representation for the fermionic determinant derived in the previous section. Let us start by performing the following transformation:

\beqn
\Psi(x) = \exp{g[\gamma_5 \Phi(x) - i \eta(x)]} ~ \chi(x) \nonumber \\
\bP(x) = \bar\chi(x) ~ \exp{g[\gamma_5 \Phi(x) + i\eta(x)]}
\label{23}
\eeqn

\beq
  \mbox{$\mathcal{D}$} \bP   \mbox{$\mathcal{D}$} \Psi = J_F [\Phi,\eta]  \mbox{$\mathcal{D}$} \bar\chi   \mbox{$\mathcal{D}$}\chi,
\label{24}
\eeq

\noindent where $\Phi$ and $\eta$ are scalar fields and $J_F [\Phi,\eta]$
is the Fujikawa Jacobian \cite{Fu} whose non-triviality is due to the
non-invariance
of the path-integral measure under chiral transformations. As it is well known, the above transformation permits to decouple the field $A_{\mu}$ from the fermionic fields if one writes

\beq
A_{\mu}(x) = \partial_{\mu}\eta(x) + \epsilon_{\mu \nu}\partial_{\nu}\Phi(x) \label{25}
\eeq
\noindent which can also be considered as a bosonic change of variables with trivial (field independent) Jacobian. As a result we find
\beq
det (i \slp - g \lnA -g' \Gamma _{\mu}C_{\mu}) = J_F[\Phi,\eta] det (i\slp-g' e^{2g \gamma_5 \Phi} \Gamma _{\mu}C_{\mu})
\label{26}
\eeq

\noindent The computation of the fermionic Jacobian requires the choice of a regularization procedure which, in turn , involves certain ambiguity (See for instance \cite{amb}). In this paper we follow the same prescription that allowed us to get sensible results when only forward scattering was considered \cite{NLT}. The result is
\beq
J_F[\Phi,\eta] = \exp{\frac{g^2}{2\pi} \int d^2x~\Phi \partial_{\mu} \partial_{\mu} \Phi}
\label{27}
\eeq

\noindent As a consequence of the above transformations the partition function is now expressed as:

\beq
Z= N'\int  \mbox{$\mathcal{D}$}\Phi  \mbox{$\mathcal{D}$}\eta \mbox{$\mathcal{D}$}C_{\mu} e^{-(S[\Phi,\eta]+S[C_{\mu}])}J_F[\Phi,\eta] \det (i\slp-g' e^{2g \gamma_5 \Phi} \Gamma _{\mu}C_{\mu})
\label{28}
\eeq

\noindent where $S[\Phi,\eta]$ arises when one inserts (\ref{25}) in $S[A_{\mu}]$
(See equation (\ref{21})).

\hs The fermionic determinant in the above expression can be analyzed in terms of a perturbative expansion. Indeed, taking $g'$ as perturbative parameter, and using the fermionic fields $\chi$ and $\bar\chi$ defined in (\ref{23}) one can write
\beq
Z_F= \sum_{n=0}^{\infty}\frac{g'^{n}}{n!}\langle \prod_{j=1}^{n} \int d^2 x_{j} \bc (x_{j})~\mbox{$ \mathsf{C} $} (x_{j}) ~\chi (x_{j}) \rangle _0
\label{29}
\eeq
where, for later convenience we have defined

\beqn
Z_F&=& \det (i\slp-g' e^{2g \gamma_5 \Phi} \Gamma _{\mu}C_{\mu}) \nonumber\\
\label{30}
\eeqn
\noindent and

\beqn
\mathsf{C}= \left( \begin{array}{cc}
C_{+} &  0 \\
0 & C_{-}
\end{array} \right)
\label{31}
\eeqn

\noindent with

\beqn
 \left \{  \begin{array}{l}
C_{+}= (C_0 + C_1) ~ e^{2g \Phi (x)} \\
C_{-}= (C_0 - C_1) ~ e^{-2g \Phi (x)}
 \end{array} \right.
\label{32}
\eeqn

\hs By carefully analyzing each term in the series we found a selection rule quite similar to the one obtained in the path-integral treatment of (1+1) massive fermions with local \cite{N} and non-local interactions \cite{KN}. Indeed, due to the fact that in (\ref{29}) $\langle~\rangle_0$ means v.e.v. with respect to free massless fermions, the v.e.v.'s corresponding to
$j = 2k+1$ are zero. Thus, we obtain
\beqn
Z_F&=&\sum_{k=0}^{\infty} \frac{(g'c \rho)^{2k}}{(k!)^2
(2\pi)^{2k}}
\int  \prod_{i=1}^{k} d^2 x_{i} d^2 y_{i} ~\nonumber\\ & \times &
\prod_{i=1}^{k} (C_0 ( x_{i}) + C_1 ( x_{i}) )(C_0 ( y_{i}) - C_1 ( y_{i}))
\nonumber\\
& \times & \exp \{2g \sum_{i=1}^{k} ( \Phi ( x_{i}) - \Phi ( y_{i})) \} \nonumber\\ & \times &
\frac{ \prod_{i>j}^{k}( (c \rho)^2 \mid  x_{i}-  x_{j}\mid  ~ \mid  y_{i}-  y_{j}\mid )^2 }
{\prod_{i,j}^{k}(c \rho \mid  x_{i}-  y_{j}\mid  )^2}
\label{33}
\eeqn
where $c\rho$ is a normal ordering parameter.

In order to obtain a bosonic description of the present problem we shall now propose the
following bosonic Lagrangian density:

\beq
\mbox{ $\mathcal{L}$ }_B= \frac{1}{2} (\partial _{\mu} \varphi )^2 +
\frac{\alpha_0}{2\beta^2} (m_{+} e^{i\beta \varphi} + m_{-} e^{ -i \beta
\varphi})
\label{34}
\eeq

\noindent with $ \beta$, $m_{+}(x)$ and $m_{-}(x)$ to be determined. The quantity $\alpha_0$ is just a constant that we include to facilitate comparison of our procedure with previous works on local bosonization \cite{massive} \cite{N}. Please notice that for $m_{+} = m_{-} = 1$ this model coincides with the well known sine-Gordon model that can be used to describe a neutral Coulomb gas. In this context $\frac{\alpha_0}{\beta^2}$ is nothing but the corresponding fugacity \cite{Samuel}.
We shall now consider the partition function
\beq
Z_B=  \int  \mbox{$\mathcal{D}$} \varphi ~\exp{- \int d^2 x  \mbox{ $\mathcal{L}$ }_B}
\label{35}
\eeq
\noindent and perform a formal expansion taking the fugacity as perturbative parameter.
It is quite straightforward to extend the analysis of each term, already performed for
$m_{+} = m_{-} = 1$, to the present case in which these objects are neither equal nor
necessarily constants. The result is

\beqn
Z_B&=& \sum_{l=1}^{\infty} \frac{1}{(l!)^2}(\frac{\alpha}{2\beta^2})^{2l} \int ( \prod_{i=1}^l d^2 x_i ~d^2 y_i) ( \prod _{i=1}^l m_{+}(x_i) ~m_{-}(y_i) )  \nonumber\\
&& \frac{ \prod_{i>j}^{l}( (c \rho)^2 \mid  x_{i}-  x_{j}\mid  ~ \mid  y_{i}-  y_{j}\mid )^{\frac{1}{2 \pi}\beta ^2} }{\prod_{i,j}^{l}(c \rho \mid  x_{i}-  y_{j}\mid  )^{ \frac{1}{2 \pi}\beta ^2}}.
\label{36}
\eeqn
\noindent where we have defined a renormalized quantity $\alpha$ (remember that there are
infrared and ultraviolet singularities involved in the correlation functions that contribute
to every order).

Comparing  this result with equation (\ref{33}), we see that both series coincide if the
following identities hold:
\beqn
\beta &=& \pm 2\sqrt{\pi} \nonumber\\
\alpha&=& g'
\label{37}
\eeqn
\noindent and

\beqn
 m_{+}(x_i)&=& \left(C_0(x_i ) + C_1(x_i )\right) ~ e^{2g \Phi (x_i )} \nonumber\\
 m_{-}(y_i )&=& \left(C_0 (y_i )- C_1 (y_i )\right) ~ e^{-2g \Phi (y_i )}
\label{38}
\eeqn
This is our second non-trivial result. We have found a bosonic representation for the
fermionic determinant (\ref{33}). This is given by (\ref{35}) together with the identities
(\ref{37}) and (\ref{38}). Let us emphasize that equations (\ref{37}) are completely
analogous to the bosonization formulae first obtained by Coleman \cite{massive} whereas
equations (\ref{38}) constitute a new result, specially connected to the present problem.

\section{Final result and next steps}
\setcounter{equation}{0}

In this paper we have presented an extension of the recently proposed NLT, which can be used
to describe a system of 1d strongly correlated particles when not only forward but also
backward and umklapp scattering is considered. We were able to write the vacuum to vacuum
functional of this model in terms of a non-trivial fermionic determinant (See eq.(\ref{33})).
 Our main achievement was to obtain a bosonic representation for this determinant
(See equations (\ref{35}),(\ref{37}) and (\ref{38})). Using this result in (\ref{28}) we
have expressed the partition function of this system  in terms of five scalars: $\Phi$,
 $\eta$, $C_0$, $C_1$ and $\varphi$. Based on our previous experience with the fs model
\cite{NLT} we know that the physical fields (the ones that describe the collective
excitations of the system) are $\Phi$ and $\eta$. The others are just auxiliary variables
that we have to integrate in order to analyze the physics of the normal modes. The
integrals in $C_0$ and $C_1$, being quadratic can be easily performed yielding:

\beqn
Z = N \int \mbox{$\mathcal{D}$} \Phi ~ \mbox{$\mathcal{D}$} \eta
~ \mbox{$\mathcal{D}$} \varphi ~ e^{-S_{eff}[\Phi,\eta,\varphi]}
\label{39}
\eeqn

\noindent with

\beqn
S_{eff}[\Phi,\eta,\varphi] & = & \frac{1}{2} \int d^2 x d^2 y (\partial _{\mu} \eta +
\epsilon_{\mu \nu} \partial _{\nu} \Phi )(x) b_{(\mu)}(x,y)(\partial _{\mu} \eta +
\epsilon_{\mu \nu} \partial _{\nu} \Phi )(y) \nonumber\\ & + &  \int d^2
x [\frac{g^2}{2 \pi}( \partial _{\mu} \Phi )^2 + \frac{1}{2} ( \partial _{\mu}
\varphi )^2] \nonumber\\ & - & \frac{1}{2} ( \frac{g'}{2 \pi})^2 \int d^2 x d^2 y
[f_{\mu}(x) ~ U_{(\mu)}(x,y) ~ f_{\mu}(y)]
\label{40}
\eeqn

\noindent where
\beqn
f_{0}(x)&=& \cosh (2 g \Phi + i \sqrt{4 \pi}\varphi)(x) \nonumber\\
f_{1}(x)&=& \sinh (2 g \Phi + i \sqrt{4 \pi}\varphi)(x)
\label{41}
\eeqn

 The analysis of the action $S_{eff}[\Phi,\eta,\varphi]$ is beyond the scope of the present
article. We are currently studying this problem by means of a saddle point computation. We hope
to report our results in the close future.

\section*{Acknowledgements}

This work was partially supported by Universidad Nacional de La Plata  and
Consejo Nacional de Investigaciones Cient\'{\i}ficas y T\'ecnicas,
CONICET (Argentina). CN thanks Igor Korepanov for his invitation
to participate in the Second International Conference on Exactly
Solvable Models, Chelyabinsk, Russia, August 1998.

\end{document}